# Photon Pair Generation in Silicon Micro-Ring Resonator with Reverse Bias Enhancement

Erman Engin,[1] Damien Bonneau,[1] Chandra M. Natarajan,[2] Alex Clark,[1] M. G. Tanner,[2] R. H. Hadfield,[2] Sanders N. Dorenbos,[3] Val Zwiller,[3] Kazuya Ohira,[4] Nobuo Suzuki,[4] Haruhiko Yoshida,[4] Norio Iizuka,[4] Mizunori Ezaki,[4] Jeremy L. O'Brien,[1] Mark G. Thompson[1*]

[1]*Centre for Quantum Photonics, H. H. Wills Physics Laboratory & Department of Electrical and Electronic Engineering, University of Bristol, Merchant Venturers Building, Bristol, UK*
[2]*Scottish Universities Physics Alliance and School of Engineering and Physical Sciences, Heriot-Watt University, Edinburgh, EH14 4AS, UK*
[3]*Kavli Institute of Nanoscience, TU Delft, The Netherlands*
[4]*Corporate Research & Development Center, Toshiba Corporation, Japan*
*Corresponding author: mark.thompson@bristol.ac.uk*

Photon sources are fundamental components for any quantum photonic technology. The ability to generate high count-rate and low-noise correlated photon pairs via spontaneous parametric down-conversion using bulk crystals has been the cornerstone of modern quantum optics. However, future practical quantum technologies will require a scalable integration approach, and waveguide-based photon sources with high-count rate and low-noise characteristics will be an essential part of chip-based quantum technologies. Here, we demonstrate photon pair generation through spontaneous four-wave mixing in a silicon micro-ring resonator, reporting a maximum coincidence-to-accidental (CAR) ratio of 602 ± 37, and a maximum photon pair generation rate of 123 MHz ± 11 KHz. To overcome free-carrier related performance degradations we have investigated reverse biased p-i-n structures, demonstrating an improvement in the pair generation rate by a factor of up to 2, with negligible impact on CAR.

Integrated quantum photonic circuits are a promising approach to realizing future quantum information processing technologies [1]. Recent demonstrations include quantum logic gates [2], small-scale factoring algorithms [3], quantum simulations [4] and entanglement generation and manipulation [5]. However, these circuits lack the crucial ability to generate quantum-states of light on-chip, and future quantum photonic circuits will require on-chip single-photon sources, single-photon detectors and compact and scalable waveguide circuits to realize applications in quantum communication, sensing, simulations and computation.

The silicon-on-insulator (SOI) material system is a particularly appealing photonic technology platform for realizing future quantum technologies as its high refractive index contrast and mature fabrication techniques provide the ability to implement compact and complex quantum circuits. In addition, the high confinement of light and the large $\chi^{(3)}$ non-linearity of silicon can be utilized for efficient photon-pair generation via spontaneous four-wave mixing (SFWM) [6]. Together with the development of near unity efficient silicon waveguide single-photon detectors [7], and the recent demonstration of quantum interference and entanglement manipulation in silicon waveguide circuits [8], silicon is a promising platform for fully integrated on-chip quantum photonic technologies.

In this paper we exploit the high $\chi^{(3)}$ nonlinearity of silicon and investigate high Q-factor micro-ring resonators to realize low-noise and high-brightness on-chip photon pair generation. SFWM for photon-pair generation has been previously demonstrated in both straight waveguides [6], and micro-ring resonators [9]. The major distinction between the two approaches is the generation of broadband photon pairs from straight waveguides, whilst micro-ring resonators can have an emission bandwidth orders of magnitude narrower. High Q-factor micro-ring resonators are also able to provide a significant field enhancement, resulting in enhanced photon-pair generation rates. However, due to the large field intensities within these structures, nonlinear optical losses associated with two-photon-absorption (TPA) induced free-carrier-absorption (FCA) can lead to degradation in the performance of the device. Here we present a low-noise and high photon pair count-rate on-chip photon source, demonstrating a two-fold performance improvement in count-rate by mitigating these parasitic nonlinear effects. A reverse-biased p-i-n diode is embedded in a silicon micro-ring resonator to remove free-carriers generated by the pump and overcome the FCA associated optical losses. To put this work into context, Table 1 presents a summary of the state-of-the-art in silicon-based photon pair sources, highlighting recent achievements in coincidence-to-accidentals (CAR) and associated photon count rates for both pulse excitation and continuous wave excitation.

Fig. 1 shows the experimental setup used. An external-cavity diode laser (ECDL) operating close to 1550 nm provides a pump beam, which is amplified by an erbium-doped fiber amplifier (EDFA) and then spectrally cleaned using a 0.5 nm bandwidth fiber Bragg grating (FBG) and a circulator. This pump beam is then injected into the chip using a 2-µm spot-size lensed fiber. The output containing the residual pump beam and also the photon pairs generated by the SFWM process are collected by another lensed fiber. The pump beam is removed using another FBG-circulator setup and the remaining photon pairs are then separated by a dense-wavelength-division-multiplexer (DWDM) with a 200 GHz channel spacing, and sent to superconducting single photon detectors (SSPD) for time of arrival analysis. We have used SSPDs mounted in a practical closed-cycle refrigerator, with maximum efficiencies of 5% and 15% at 1550 nm [10]. The detection correlation measurement is performed by the time interval analyzer (TIA).

Table 1. Selection of silicon-based sources in the literature, showing maximum values for CAR and associated photon count rates for pulsed excitation and continuous wave excitation for both straight waveguide sources and ring resonator sources.

| Ref | Structure | CAR | Photon count rate |
|---|---|---|---|
| **Pulsed Excitation** | | | |
| [6] | 9.11mm waveguide | 25 | 2.5MHz |
| [11] | 1.15cm waveguide | 320 | 0.1MHz |
| [12] | 1.09cm waveguide | 50 | 0.5MHz |
| **CW Excitation** | | | |
| [9] | 1.13cm waveguide | 11 | 9MHz |
| [9] | 43µm Si ring resonator | 30 | ~0.5MHz |
| this work | 73µm Si ring resonator | 600 | 0.8MHz |

The device is fabricated using an SOI layer structure with a 220 nm silicon layer thickness. The etch depth is 170 nm and the width of the waveguide is 450 nm. The chip is further coated with an oxide layer of thickness 90 nm. The length of the device is 2 mm having spot size converters on both the input and the output to efficiently couple the light. The resonator is a racetrack with 10 µm bend radius and 5 µm straight sections. The coupling losses between the facets and the fiber lenses are estimated to be 3 dB per facet, with a total device insertion loss of ~10 dB. On resonance the device exhibits a 13 dB extinction ratio, and has a Q factor of 37,500. Throughout the experiment we have used TE polarized light.

The resonance wavelength of the micro-ring is controlled via temperature tuning, and when the pump wavelength and the ring resonance are not aligned, we observe photon pair generation from only the 2-mm long straight bus waveguide, as no pump power is coupled to the ring. When the ring and pump are resonant, there is strong coupling to the ring and due to the large field intensity inside of the ring (of length ~70 µm) there is enhanced photon-pair generation. At high pump powers, TPA of the pump occurs and the resulting free-carrier absorption (FCA), free-carrier dispersion (FCD) and thermo-optical effects give rise to a shift of the ring resonance to a higher wavelength, and an associated reduction in the Q-factor due to the nonlinear losses within the ring.

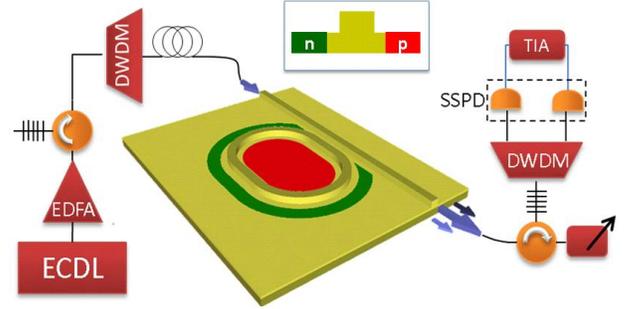

Fig.1 A tunable external cavity diode laser (ECDL) is amplified and the noise is removed using a fibre Bragg grating (FBG) with circulator and a dense wavelength division multiplexer (DWDM). The light is coupled in an out of the device using lensed fibers. The output pump and generated photons are separated with another FBG and DWDM. The channels of the DWDM are input into the two channels of the superconducting single photon detector (SSPD) and the response is recorded using a time interval analyzer (TIA). The device is a ring resonator with a 10 µm radius and 5 µm coupling region. The inset shows the cross section of the waveguide and p-i-n junction.

Fig. 2a presents the relation between the intrinsic photon pair generation rate and the pump power inside the device. When the ring is off-resonance, the pair generation rate is observed to have a quadratic relationship with the input power, whilst the pair generate rate when the pump is on-resonance are fitted with the following equation derived from the work presented in [13], which accounts for the nonlinear relation between the power inside the ring and the injected pump power.

$$P_{in} = \sqrt{CC/(at^4)}(1 - r\tau(1 + CC\beta/a)^{-1/4})^2 \quad (1)$$

In equation (1) $CC$ is the number of coincidence counts (pairs) measured per second, $\alpha$ is the nonlinear coefficient which links $CC$ and the power inside the ring, $r$ and $t$ are the reflection and transmission coefficients of the waveguide coupled to the ring, $\beta$ is the nonlinear coefficient accounting for free carrier absorption, and $1-\tau$ is the amplitude reduction factor per round trip. The power injected to the ring is calculated from the input power accounting for the facet loss.

By taking into account the overall losses of the two collection channels including the detector efficiencies (calculated as -27 dB and -34 dB from measurement of the singles and coincidence counts - see methods section), the intrinsic pair generation rate coefficients are calculated to be 45 KHz.mW$^{-2}$ for off-resonance and 3.1 MHz.mW$^{-2}$ for on-resonance. For a fixed injected power of 5.3 mW, a comparison of the off- and on-resonance cases gives an enhancement in count rate from 998 kHz to 88 MHz — a 88-fold increase.

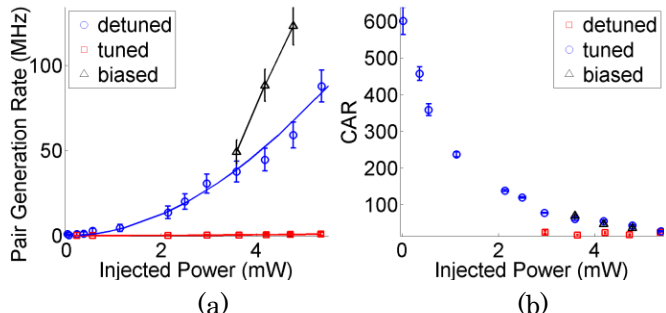

Fig. 2(a) The on-chip photon pair generation rate versus the power input to the device for off-resonance, on-resonance and 8 V reverse bias applied. Count rate is taken as the integral over the FWHM of coincidence histogram. (b) The CAR plotted for these three cases and taken also as the integral over the FWHM - see methods section for more details.

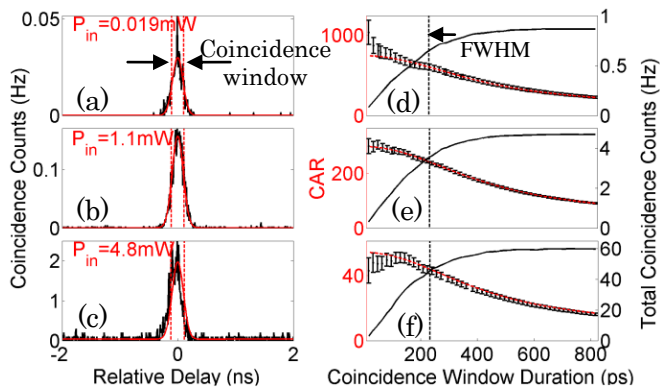

Fig.3 (a-c) Coincidence histograms acquired by time interval analysis, with Gaussian fit (red), for different input powers. (d-f) CAR (error bars, left axis) and the total coincident counts (solid, right axis) vs. coincidence window size corresponding to the histograms in (a-c). The dashed red curves in (d-f) is the CAR computed from the Gaussian fit.

It is well known that the large field enhancement of high Q-factor silicon resonators give rise to strong TPA even at relatively low pump powers, generating free-carriers and introducing additional losses which degrade the device performance [14]. In order to mitigate these effects we have incorporated a p-i-n junction surrounding the ring, and by applying a reverse-bias voltage it is possible to sweep out the free carriers generated inside the ring. This approach has previously been shown to reduce the free carrier life-time and suppress free-carrier related effects, leading for example to an enhanced Raman lasing in silicon [14]. A reverse-bias voltage of 8 V was found to give the best enhancement for the input powers investigated in this work, and Fig. 2a plots the generation rate for the reverse bias case (triangles), showing an enhancement of up to 2.1 times that of the unbiased case (dots) – for 4.8 mW injection power – with a maximum count rate of 123 MHz achieved (compared with 59MHz for the un-biased case).

The photon-pair generation rates presented in Fig. 2a are calculated by acquiring histograms of the coincidence counts for different arrival times of the signal and idler photons (with 8 ps intervals), and then integrating over a coincidence window duration ($\tau$) corresponding to the full-width-half-maximum (FWHM) of the coincidence histogram. Fig. 3a-c presents examples of three of these coincidence histograms for three different input powers (0.019, 1.1 and 4.8 mW), for the case of an un-biased on-resonance device. In Fig. 3d-f, the black solid lines show the total coincidence counts as a function of the coincidence window duration, which tends to a fixed value for coincidence window durations much greater than the FWHM of the coincidence histogram.

To calculate the coincidence-to-accidental ratio (CAR), the accidental counts are measured far from the coincidence peak, and Fig. 3d-f plots both the measured CAR values (points with error bars) and CAR values calculated from the Gaussian fit to the coincidence histogram (dashed line) - see methods for more details. It can be observed that as the coincidence window duration approaches zero, the CAR values calculated from the Gaussian fit tend to a single value. This value represents that maximum obtainable CAR value from this particular experiment, taking into account detector noise, pump leakage, Raman noise, higher order multi-photon terms, etc. The total coincidence count rate on the other hand tends to a maximum value as the coincidence window duration tends to infinity (and approaches zero as the coincidence window duration tends to zero). Thus to define single values for the photon pair generation rate and CAR, it is necessary to chose a particular coincidence window duration. Typically, the pair generation rates and CAR values quoted in the literature use a range of different and some-what arbitrarily chosen values of the coincidence window. To define the single values for pair generation rate and CAR presented in Fig. 2, we have use the FWHM of the Gaussian fit to the coincidence histogram as coincidence window duration (which is ~230ps in all cases and is defined by the jitter of the measurement setup). This enables a useful estimate of the pair generation rate and CAR values by taking into account a significant number of all the coincidence events, yielding values that are close to the maximum obtainable pair generation rates (given as $\tau \rightarrow$ infinity) and the maximum obtainable CAR (given as $\tau \rightarrow 0$). Thus 4provides a useful benchmark for comparison with previous experimental results (see Table 1). Table 2 presents a summary of the photon-pair generation rate and associated CAR for two different coincidence windows; one covering FWHM (~230 ps) and one covering ±3 standard deviations (~580ps) of the Gaussian fit to the coincidence histogram.

Table 2. Tabulate results of the CAR and pair generation rate for coincidence window durations covering FWHM and ±3 standard deviations ($\sigma$) of the Gaussian fit to the coincidence histogram.

| Injection power (mW) | CAR (FWHM) | CAR (±3$\sigma$) | Pair generation rate (MHz) | Pair generation rate (±3$\sigma$) (MHz) |
|---|---|---|---|---|
| With 8V reverse bias | | | | |
| 4.8 | 37 ± 1.3 | 19 ± 0.6 | 123 | 161 |
| No reverse bias | | | | |
| 4.8 | 43 ± 2.2 | 23 ± 1 | 59 | 75 |
| 1.1 | 359 ± 16 | 125 ± 3 | 2.7 | 6 |
| 0.019 | 602 ± 37 | 315 ± 18 | 0.827 | 1 |

A maximum CAR value of 602 ± 37 is measured for an injection pump power of 0.019 mW, corresponding to photon-pair generation rate of 827 KHz. The maximum photon pair generation rate of 123 MHz is achieved for an injection pump power of 4.8 mW and by applying reverse bias voltage across the p-i-n structure to mitigate the parasitic effects of free-carrier absorption (with an associated CAR value of 43 ± 2.2). These results highlight the potential of silicon as a promising photon source for photonic quantum information processing, with properties of low noise and high generation rate. It is interesting to note from comparing the CAR values shown in Fig. 2b, we see that for the on-resonance conditions, the reverse-biased and un-biased case give similar values, whereas for off-resonance the CAR is much lower. This likely due to noise from Raman scattering in optical fibers of the experimental setup which deteriorates the CAR for low photon-pair generation rates in the detuned case.

Currently the filtering of the pump, signal and idler photons are achieved using fiber optic components, but with improved architectures, on-chip filtering could also be performed by further taking advantage of resonant structures. The ring cavity also has the advantage of generating photon pairs having low spectral entanglement [15] when pumped with a pulsed source having a bandwidth greater than the resonance width. This will enable such ring source to be used as a high purity single photon sources heralded on the measurement of one of the photon pairs.

The high CAR achieved makes those sources a good candidate for low noise multiplexed single photon sources [16] which could in turn be used in a scalable manner to seed a linear optical quantum circuit.

The on-chip photon pair source reported in this paper is immediately compatible with already demonstrated silicon quantum circuits [8] and high-efficiency waveguide integrated superconducting single-photon detectors [7]. It is now therefore possible to realize in a single material system all the major components required for on-chip linear optics quantum information processing; low noise single photon sources, compact waveguide circuits and efficient single photon detectors. The next major challenge is to integrate these components into a single unified technology platform, making the silicon-on-insulator material system a promising approach for future quantum photonic technologies for applications in communication, sensing, simulation and computation.

Methods

**Calculation of pair generation rate:** In order to infer the pair generation rates on-chip we have to account for the collection efficiencies of the detection channels which includes all the losses that photons undergo from the chip to the detector and the efficiency of the detector. The collection efficiencies of the two channels and the single and the coincidence counts are related to each other by the following set of formulas.

$$C_1 = \eta_1 \beta P^2 + n_1 P + dc_1 \qquad (2)$$
$$C_2 = \eta_2 \beta P^2 + n_2 P + dc_2 \qquad (3)$$
$$CC = \eta_1 \eta_2 \beta P^2 + acc \qquad (4)$$

where $C_1$ and $C_2$ are the single photon counts, $\eta_1$, $\eta_2$ are the collection efficiencies of detection channels 1 and 2 respectively, $CC$ is the coincidence count, $P$ is the input power and $\beta$ is the pair generation rate. The coefficients $n_1$, $n_2$ are associated with the noise terms linear with input power (such as Raman noise from the fibers), $dc_1$ and $dc_2$ are the dark counts of the two detectors and $acc$ is the accidental counts. The quadratic coefficients of the input power, $P$ in all three equations can be extracted from polynomial fits to the measurement data. From these quadratic coefficients it is possible to extract the values for $\eta_1$, $\eta_2$ and $\beta$. We have integrated the full coincidence histogram peak to obtain the $CC$ values in equation (4) for low input powers (<2.1mW) in the tuned case. The channel collection efficiencies are found to be -27 dB and -34 dB for channel 1 and channel 2 respectively, and takes into account all losses within the system (including device coupling losses). These values are used in the calculation of the pair generation rate throughout the experiments.

**Calculation of the CAR:** The CAR is calculated by acquiring a histogram of the coincidence counts within an 8 ps detection time interval between signal and idler photons.

Fig. 3a-c shows examples of three coincidence histograms for three different input powers. A Gaussian curve is fitted to the histograms and the *CC* is defined as the sum of the counts within the given coincidence window depicted in Fig. 3a-c. The accidental counts are calculated far outside the peak. The CAR value and associated *CC* depends greatly on the duration of the defined coincidence time window. Fig. 3d-f plots this dependence of the CAR and *CC* on the duration of the chosen coincidence window corresponding to the histograms in Fig. 3a-c respectively. The dashed red curves in Fig. 3d-f correspond to CAR values when the fitted Gaussian curve shown with solid red curves in Fig. 3a-c are used instead of the measured count values. The values of CAR has a decreasing trend as the coincidence window gets wider while the *CC* saturates typically after an integration window width of 2xFWHM. To define a single value of CAR and photon pair generation rate (as presented in Fig. 2), we have used the coincidence window corresponding to the FWHM of the Gaussian fit.

# References

1. Thompson, M. G., Politi, A., Matthews, J. C. F. & O'Brien, J. L. Integrated waveguide circuits for optical quantum computing. Circuits, Devices & Systems, IET **5**, 94–102 (2011).
2. Politi, A., Cryan, M. J., Rarity, J. G., Yu, S. & O'Brien, J. L. Silica-on-silicon waveguide quantum circuits. Science **320**, 646–9 (2008).
3. Politi, A., Matthews, J. C. F. & O'Brien, J. L. Shor's quantum factoring algorithm on a photonic chip. Science **325**, 1221 (2009).
4. Peruzzo, A. et al. Quantum walks of correlated photons. Science **329**, 1500 (2010).
5. Shadbolt, P. J. et al. Generating, manipulating and measuring entanglement and mixture with a reconfigurable photonic circuit. Nature Photonics **6**, 45–49 (2011).
6. Sharping, J. E. et al. Generation of correlated photons in nanoscale silicon waveguides. Opt. Express **14**, 12388–12393 (2006).
7. Pernice, W. et al. High speed and high efficiency travelling wave single-photon detectors embedded in nanophotonic circuits. arXiv **1108.5299**, (2011).
8. Bonneau, D. et al. Quantum interference and manipulation of entanglement in silicon wire waveguide quantum circuits. New Journal of Physics **14**, 045003 (2012).
9. Clemmen, S. et al. Continuous wave photon pair generation in silicon a-on-insulator waveguides and ring resonators. Opt. Express **17**, 16558–16570 (2009).
10. Tanner, M. G. et al. Enhanced telecom wavelength single-photon detection with NbTiN superconducting nanowires on oxidized silicon. Applied Physics Letters **96**, 221109 (2010).
11. Rukhlenko, I. D., Premaratne, M. & Agrawal, G. P. Analytical study of optical bistability in silicon ring resonators. Optics letters **35**, 55–57 (2010).
12. Rong, H. et al. An all-silicon Raman laser. Nature **433**, 292–4 (2005).
13. Helt, L. G., Yang, Z., Liscidini, M. & Sipe, J. E. Spontaneous four-wave mixing in microring resonators. Optics Letters **35**, 3006 (2010).
14. Migdall, A., Branning, D. & Castelletto, S. Tailoring single-photon and multiphoton probabilities of a single-photon on-demand source. Physical Review A **66**, (2002).
15. Harada, K. I. et al. Frequency and polarization characteristics of correlated photon-pair generation using a silicon wire waveguide. Selected Topics in Quantum Electronics, IEEE Journal of **16**, 325–331 (2010).
16. Takesue, H. et al. Entanglement generation using silicon wire waveguide. Applied Physics Letters **91**, 201108–201108–3 (2007).

# Acknowledgments

We acknowledge John Rarity, Graham Marshall and Josh Silverstone for useful discussions. This work was supported by EPSRC, ERC, the Centre for Nanoscience and Quantum Information (NSQI) and the European FP7 project QUANTIP. MGT acknowledges support from the Toshiba Research Fellowship scheme. JLO'B acknowledges a Royal Society Wolfson Merit Award. RHH acknowledges a Royal Society University Research Fellowship.